\documentclass[twocolumn,showpacs,aps,pre,nofootinbib]{revtex4}
\usepackage{hyperref}
\usepackage{graphicx}
\usepackage{amsmath,amsfonts,amsthm}
\usepackage{color}

\usepackage{grffile}
\usepackage{subfigure}
\usepackage[normalem]{ulem}

\usepackage{tikz}
\usetikzlibrary{arrows}
\usetikzlibrary{calc}
\usetikzlibrary{patterns}
\usetikzlibrary{scopes}

\newcommand{\beq}{\begin{equation}}
\newcommand{\eeq}{\end{equation}}

\newcommand{\<}{\langle}
\renewcommand{\>}{\rangle}

\DeclareMathOperator*{\argmax}{arg\,max}

\numberwithin{equation}{section}

\begin{document}
\title{Weighting of topologically different interactions in a model of two-dimensional polymer collapse}
\author{Andrea Bedini} \email{abedini@ms.unimelb.edu.au}
\affiliation{Department of Mathematics and Statistics, The University
  of Melbourne, 3010, Australia}
\author{Aleksander L.\ Owczarek} \email{owczarek@unimelb.edu.au}
\affiliation{Department of Mathematics and Statistics, The University
  of Melbourne, 3010, Australia}
\author{Thomas Prellberg} \email{t.prellberg@qmul.ac.uk}
\affiliation{School of Mathematical Sciences, Queen Mary University of
  London, Mile End Road, London, E1 4NS, United Kingdom}

\begin{abstract}
  We study by computer simulation a recently introduced generalised
  model of self-interacting self-avoiding trails on the square lattice
  that distinguishes two topologically different types of
  self-interaction: namely \emph{crossings} where the trail passes
  across itself and \emph{collisions} where the lattice path visits
  the same site without crossing. This model generalises the canonical
  interacting self-avoiding trail model of polymer collapse which has
  a strongly divergent specific heat at its transition point. We
  confirm the recent prediction that the asymmetry does not affect the
  universality class for a range of asymmetry. Certainly, where the
  weighting of collisions outweighs that of crossings this is well
  supported numerically. When crossings are weighted heavily relative
  to collisions the collapse transition reverts to the canonical
  $\theta$-point-like behaviour found in interacting self-avoiding
  walks.
\end{abstract}



\maketitle

\section{Introduction}
\label{sec:introduction}

The collapse transition of a polymer in a dilute solution has been a
continuing focus of study in lattice statistical mechanics for decades
\cite{gennes1975a-a,gennes1979a-a}. This transition describes the
change in the scaling of the polymer with length that occurs as the
temperature is lowered. At high temperatures the radius of gyration of
polymer scales in a way swollen relative to a random walk: this is
known as the excluded volume effect. At low temperatures a polymer
condenses into dense, usually disordered, globule, with a much smaller
radius of gyration.  The interest in this phase transition has
occurred both because of the motivation of physical systems but also
because of the study of integrable cases
\cite{nienhuis1982a-a,warnaar1992b-a} of lattice models have proved
especially fruitful in two dimensions.  While the canonical lattice
model of the configurations of a polymer in solution has been the
self-avoiding walk (SAW), where a random walk on a lattice is not
allowed to visit a lattice site more than once, an alternative has
been to use bond-avoiding walks, or a self-avoiding trail. A
self-avoiding trail (SAT) is a lattice walk configuration where the
excluded volume is obtained by preventing the walk from visiting the
same bond, rather than the same site, more than once. The model of SAT was used
initially to model polymers with loops \cite{malakis1976a-a} but has
subsequently occurred in integrable loop models in two dimensions
\cite{warnaar1992b-a}.  A model of collapsing polymers can be
constructed starting from self-avoiding trails, known as interacting
self-avoiding trails (ISAT). Here energies are associated with
multiply-visited sites, and by favouring configurations with many such
sites a collapse transition can be initiated.

Owczarek and Prellberg studied numerically the ISAT collapse on the
square lattice by two different approaches
\cite{owczarek1995a-:a,owczarek2007c-:a} and in either case found a
strong continuous transition with specific heat exponent $\alpha =
0.81(3)$. Recently, on the triangular lattice Doukas \textit{et al.\
}\cite{doukas2010a-:a} found that by changing the weighting of doubly
and triply visited sites a first-order transition can ensue or
alternatively, depending on the ratio of these weightings, a weaker
second-order transition that mimics the collapse found in the
canonical interacting self-avoiding walk (ISAW) model (also known as
the $\theta$-point). They also found that the low temperature phase
becomes fully dense rather than globular, depending on the choice of
parameters.

\begin{figure}[t!]
  \centering
  \begin{tikzpicture}
    \draw[help lines] (-1,0) grid (6,3); \fill (0,0) circle (2pt);
    \draw[->,very thick,rounded corners=5pt] (0,0) -- ++(0,2) --
    ++(1,0) node [below left=2pt] {$\tau_c$} -- ++(0,-1) -- ++(1,0) node [below left=2pt] {$\tau_c$} --
    ++(0,-1) -- ++(1,0) -- ++(0,1) node [below right=2pt] {$\tau_x$} -- ++(0,2) -- ++(2,0) -- ++(0,-2) -- ++(-3,0)
    -- ++(0,1) -- ++(-1,0) -- ++(0,1) ;
  \end{tikzpicture}
  \caption{An example of AISAT configuration with one crossing ($m_x = 1$) and two collisions ($m_c=2$), associated 
   with the Boltzmann weights $\tau_x$ and $\tau_c$, respectively. The total number of doubly visited sites is $m_2 = 3$.  }
  \label{fig:aisat}
\end{figure}
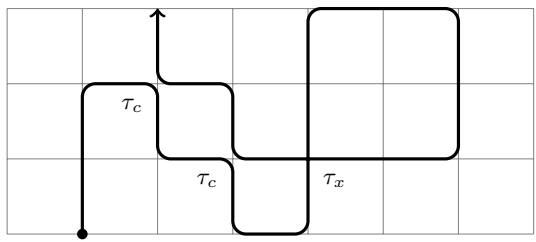

Recently, Foster \cite{foster2011a-a} generalised the ISAT model on the square lattice by differentiating the type of doubly visited sites on the square lattice: a doubly visited site can be visited twice with the trail passing through itself, that is \emph{crossing}, at the site, or alternatively as a result of two bends in the trail so that the trail ``touches" or ``collides": see Fig.~\ref{fig:aisat}.  \label{sec:model}

We shall refer to this model as the asymmetric ISAT model (AISAT). The
study \cite{foster2011a-a} using transfer matrices and the
phenomenological renormalisation group of the AISAT predicted that the
universality class of the symmetric case extended to the asymmetric
case. However, when crossings sufficiently dominate over collisions
the results suggested the appearance of a first-order transition.

In this work we use Monte Carlo simulation to explore this AISAT model
and the predictions of Foster \cite{foster2011a-a}. We also explore
the low temperature phase of the model and find that it is fully dense
for a range of asymmetry including the symmetric case.

\section{ISAT}
\label{sec:sym-isat}

The model of interacting trails on the square lattice is defined as
follows. Consider the ensemble $\mathcal T_n$ of self-avoiding trails
(SAT) of length $n$, that is, of all lattice paths of $n$ steps that
can be formed on the square lattice such that they never visit the
same bond more than once.
Given a SAT $\psi_n \in \mathcal T_n$, we associate an energy
$-\varepsilon_t$ with each doubly visited site and denote by
$m_j(\psi_n)$ the number sites visited $j$ times by $\psi_n$. We have
$n = m_1 + 2 m_2$.  The probability of $\psi_n$ is given by
\begin{equation}
	\frac{e^{\beta \varepsilon_t m_2(\psi_n)}}{Z^{ISAT}_n(T)},
\end{equation}
where we define the Boltzmann weight $\omega_t = \exp(\beta
\varepsilon_t)$ and $\beta$ is the inverse temperature $1/k_BT$. The
partition function of the ISAT model is given by
\begin{equation}
	Z^{ISAT}_n(T) = \sum_{\psi_n\in \mathcal T_n}\ \omega_t^{m_2(\psi_n)}.
\end{equation}
The finite-length reduced free energy is
\begin{equation}
	\kappa_n(T) = \frac{1}{n} \log\ Z_n(T)
\end{equation}
and the thermodynamic limit is obtained by taking the limit of large $n$, i.e.,
\begin{equation}
	\kappa(T) = \lim_{n \to \infty} \kappa_n(T).
\end{equation}
It is expected that there is a collapse phase transition at a
temperature $T_c$ characterized by a non-analyticity in $\kappa(T)$.

The collapse transition can be characterized via a change in the
scaling of the size of the polymer with temperature. It is expected
that some measure of the size, such as the radius of gyration or the
mean squared distance of a monomer from the end points, $R_n^2(T)$,
scales at fixed temperature as
\begin{equation}
  R_n^2(T) \sim A n^{2\nu}
\end{equation}
with some exponent $\nu$. At high temperatures the polymer is swollen
and in two dimensions it is accepted that $\nu=3/4$
\cite{nienhuis1982a-a}. At low temperatures the polymer becomes dense
in space, though not space filling, and the exponent is $\nu=1/2$. The
collapse phase transition is expected to take place at some
temperature $T_c$. If the transition is second-order, the scaling at
$T_c$ of the size is intermediate between the high and low temperature
forms. In the thermodynamic limit there is expected to be a
singularity in the free energy, which can be seen in its second
derivative (the specific heat). Denoting the (intensive) finite length
specific heat \emph{per monomer} by $c_n(T)$, the thermodynamic limit
is given by the long length limit as
\begin{equation}
  C(T) = \lim_{n\rightarrow\infty} c_n(T)\;.
\end{equation}
One expects that the singular part of the specific heat behaves as
\begin{equation}
  C(T) \sim B |T_c -T|^{-\alpha}\;,
\end{equation}
where $\alpha<1$ for a second-order phase transition.  The singular
part of the thermodynamic limit internal energy behaves as
\begin{equation}
  U(T) \sim B |T_c -T|^{1-\alpha}
\end{equation}
if the transition is second-order, and there is a jump in the internal
energy if the transition is first-order (an effective value of
$\alpha=1$).

Moreover one expects crossover scaling forms \cite{brak1993a-:a} to
apply around this temperature, so that
\begin{equation}
  c_n(T) \sim n^{\alpha\phi} \; {\cal C}((T - T_c)n^\phi)
\label{spec-heat-scaling}
\end{equation}
with  $0<\phi < 1$ if the transition is second-order, and 
\begin{equation}
  c_n(T) \sim n \; {\cal C}((T - T_c)n)
\end{equation}
if the transition is first-order. From \cite{brak1993a-:a} we point
out that the exponents $\alpha$ and $\phi$ are related via
\begin{equation}
  2-\alpha = \frac{1}{\phi}\;.
\end{equation}
Important for numerical estimation is the use of
equation~(\ref{spec-heat-scaling}) at the peak value of the specific
heat given by $y^{peak}= (T - T_c)n^\phi$ so that
\begin{equation}
  c^{peak}_n(T) \sim   {\cal C}^{peak} \; n^{\alpha\phi}
\label{spec-heat-peak-scaling}
\end{equation}
where ${\cal C}^{peak}= {\cal C}(y^{peak})$ is a constant.

A previous study \cite{owczarek2007c-:a} of ISAT model on the square
lattice has shown that there is a collapse transition with a strongly
divergent specific heat, with
\begin{equation}
  \alpha\phi =0.68(5)
\label{isat-exponent}
\end{equation}
and so the individual exponents have been estimated as
\begin{equation}
  \phi =0.84(3)\quad \mbox{ and } \quad \alpha=0.81(3)\;. 
\end{equation}
At $T=T_c$ it was  predicted \cite{owczarek1995a-:a} that
\begin{equation}
  R_n^2(T) \sim A n\left(\log n\right)^2\;.
\end{equation}

\section{Asymmetric ISAT model}

The Asymmetric ISAT (AISAT) model can be defined as follows.  Consider
the set of bond-avoiding paths $\mathcal T_n$ as defined in the
previous section. Given a SAT $\psi_n \in \mathcal T_n$, we associate
an energy with each doubly visited site,
as in ISAT, but we make a distinction between whether the trail crosses itself
or not. We will call the former \emph{crossings} and the latter
\emph{collisions} with associated energies $-\varepsilon_x$ and
$-\varepsilon_c$, respectively. For each configuration $\psi_n \in
\mathcal T_n$ we count the number $m_x(\psi_n)$ of crossings and
$m_c(\psi_n)$ of collisions, see Fig.~\ref{fig:aisat}. Note that
the total number of doubly visited sites is $m_2 = m_c + m_x$. We 
associate with each configuration a Boltzmann weight $\tau_x^{m_x(\psi_n)}
\tau_c^{m_c(\psi_n)}$ where $\tau_{x} = \exp(\beta
\varepsilon_{x})$, $\tau_{c} = \exp(\beta
\varepsilon_{c})$ and $\beta$ is the inverse temperature $1/k_B
T$. The partition function of the AISAT model is the given by
\begin{equation}
  Z_n(\tau_x, \tau_c) = \sum_{\psi_n\in\mathcal T_n}\
  \tau_x^{m_x(\psi_n)} \tau_c^{m_c(\psi_n)}
  . 
\end{equation}
The probability of a configuration $\psi_n$ is then
\begin{equation}
  p(\psi_n; \tau_x, \tau_c) = \frac{ \tau_x^{m_x(\psi_n)}
    \tau_c^{m_c(\psi_n)} }{ Z_n(\tau_x, \tau_c) }
  .
\end{equation}
In line with Foster \cite{foster2011a-a} let us define the variables
\begin{equation}
  x= \frac{\tau_x}{\tau_c}
\end{equation}
and
\begin{equation}
  r =\frac{x}{1+x} = \frac{\tau_x}{\tau_x+\tau_c}
  .
\end{equation}
When we set $\tau_x = \tau_c$ ($x=1, r=1/2$) the model reduces to the ISAT model, in
which crossings and collisions are given the same weight. On the other
end, if we set $\tau_x = 0$ ($x=r=0$) configurations with crossings are excluded
and our model reduces to the VISAW model \cite{Blote:1998ib}.

The average of any quantity $Q$ over the ensemble set of path
$\mathcal T_n$ is given generically by
\begin{equation}
  \langle Q \rangle(n; \tau_x, \tau_c) = \sum_{\psi_n\in\mathcal
    T_n} Q(\psi_n) \, p(\psi_n; \tau_x, \tau_c)
  .
\end{equation}
In particular, we can define the average number of crossings and
collisions per site and their respective fluctuations as
\begin{align}
  u_{x} &= \frac{ \langle m_{x} \rangle }{n}\;,
  &
  c_{x} &= \frac{ \langle m^2_{x} \rangle - \langle m_{x}
    \rangle^2 }{n}\;,\\
  u_{c} &= \frac{ \langle m_{c} \rangle }{n}\;,
  &
  c_{c} &= \frac{ \langle m^2_{c} \rangle - \langle m_{c}
    \rangle^2 }{n}
  .
\end{align}
An important quantity for what follows is the proportion of the sites
on the trail that are at lattice sites which are not doubly occupied:
\begin{equation}
  p_n = 1 - \frac2n \left(\< m_c \> + \< m_x \> \right)
  .
\end{equation}
Foster \cite{foster2011a-a} predicted that the universality class of
symmetric ISAT at $r=1/2$ extends to other values of $r$ and further
that there may be a change to a first-order transition for large values of $r$.

\section{Numerical results}
\label{sec:results}

We began by simulating the full two parameter space by using the
flatPERM algorithm \cite{prellberg2004a-a}. FlatPERM outputs an
estimate $W_{n,\mathbf{k}}$ of the total weight of the walks of length
$n$ at fixed values of some vector of quantities
$\mathbf{k}=(k_1,k_2,\dotsc,k_{\ell})$. From the total weight one can
access physical quantities over a broad range of temperatures through
a simple weighted average, e.g.
\begin{align}
  \< \mathcal O \>_n(\tau) = \frac{\sum_{\mathbf{k}} \mathcal O_{n,\mathbf{k}}\,
   \left(\prod_j \tau_j^{k_j}\right) \, W_{n,\mathbf{k}}}{\sum_\mathbf{k} \left(\prod_j \tau_j^{k_j}\right) \, W_{n,\mathbf{k}}}.
\end{align}
The quantities $k_j$ may be any subset of the physical parameters of
the model. In our case we begin by using $k_1=m_x$ and $k_2=m_c$.

\begin{figure}[t!] \centering
{\includegraphics{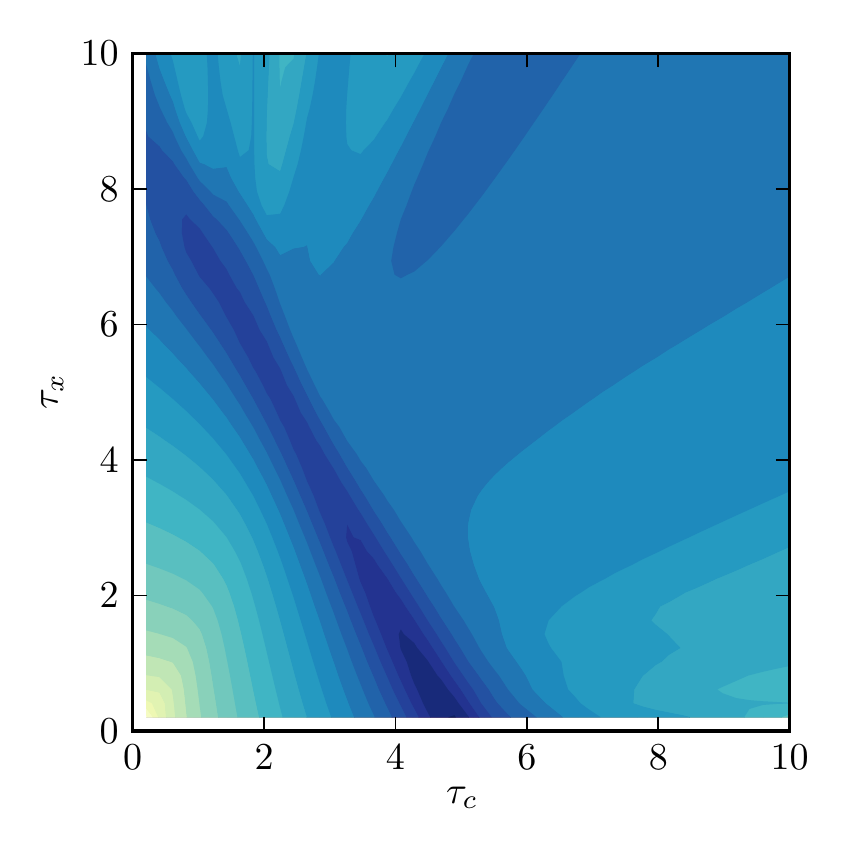}}
\caption{Density plot of the logarithm of the largest
eigenvalue $\lambda_{max}$ of the matrix of second derivatives of
the free energy with respect to $\tau_x$ and $\tau_c$ at
length $500$.}
\label{fig:eigenvalue_plot}
\end{figure}

We have simulated AISAT using the full two-parameter flatPERM
algorithm up to length $n = 500$, with $10^5$ iterations, collecting
$1.2 \cdot 10^{10}$ samples at the maximum length.  Following
\cite{prellberg2004a-a}, we also measured the number of samples
adjusted by the number of their independent growth steps (``effective
samples'') $S^{eff} \simeq 3.9 \cdot 10^8$ at the maximum length. To
obtain a landscape of possible phase transitions we plot the largest
eigenvalue of the matrix of second derivatives of the free energy with
respect to $\tau_x$ and $\tau_c$ at length $n=500$ in
Fig.~\ref{fig:eigenvalue_plot}.

We notice that there is a strong peak in the fluctuations running in a
line from 
$\tau_c \approx 5$ when $\tau_x=0$ ($r=0$)
to 
$\tau_x \approx 8$ when $\tau_c=0$ ($r=1$). It is interesting to observe that
the peak in the fluctuations becomes weaker as $r$ increases from $0$
to $1$.

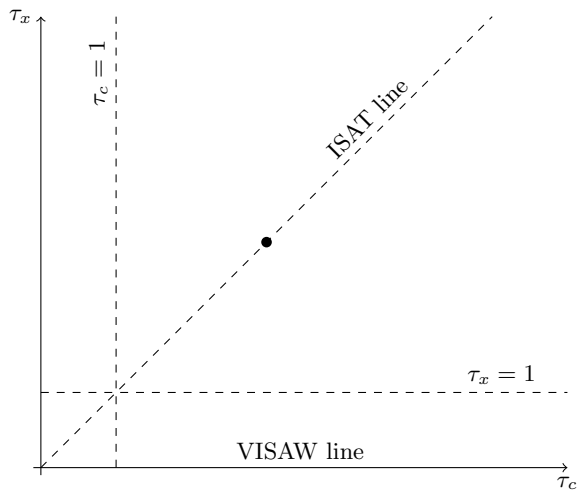
\begin{figure}[t!]
  \centering 
  \begin{tikzpicture}
    \draw[->] (-0.1,0) -- (7,0) node[below] {$\tau_c$}
    node [midway,above] {VISAW line};
    \draw[->] (0,-0.1) -- (0,6) node[left] {$\tau_x$};
    \draw[dashed] (0,0) -- (6,6) node[near end,above,sloped] {ISAT line};
    \draw[dashed] (1,0) -- (1,6) node[very near end,above,sloped] {$\tau_c = 1$};
    \draw[dashed] (0,1) -- (7,1) node[very near end,above] {$\tau_x =
      1$};
    \fill (3,3) circle (2pt);
  \end{tikzpicture}
  \caption{Schematic diagram of the AISAT parameter space. The black
    dot on the diagonal line indicates the ISAT critical point. We have
    simulated along the dashed lines, as well as the VISAW line.}
  \label{fig:phase-space}
\end{figure}

Next, we have simulated four different one parameter slices of the AISAT
model all up to length $n = 1024$. Their location, as indicated in Fig.~\ref{fig:phase-space},
is as follows.

\begin{itemize}
\item $\tau_x = 0$ (the VISAW model). With $S \simeq 1.8
  \cdot 10^7$ iterations, collecting $4.5 \cdot 10^{10}$ samples at
  the maximum length (corresponding to $S^{eff} \simeq 5.2 \cdot
  10^8$).

\item $\tau_x = 1$ (the {\it colliding} model). With $7.8 \cdot 10^6$
  iterations, collecting $2.6 \cdot 10^{10}$ samples at the maximum
  length (corresponding to $S^{eff} \simeq 4.1 \cdot 10^8$).

\item $\tau_x = \tau_c$ (the symmetric ISAT model). With
  $S \simeq 4 \cdot 10^6$ iterations, collecting $7.6 \cdot 10^9$
  samples at the maximum length (corresponding to $S^{eff} \simeq 5.8
  \cdot 10^8$).

\item $\tau_c = 1$ (the {\it crossing} model). With $S \simeq 7.7 \cdot
  10^6$ iterations, collecting $2.7 \cdot 10^{10}$ samples at the
  maximum length (corresponding to $S^{eff} \simeq 2.7 \cdot 10^8$).
\end{itemize}

\subsection{Specfic heat}
\label{sec:specific-heat}

\begin{figure*}
  \centering
  \subfigure[The {\it crossing} model, $\tau_c = 1$]{\includegraphics[scale=0.9]{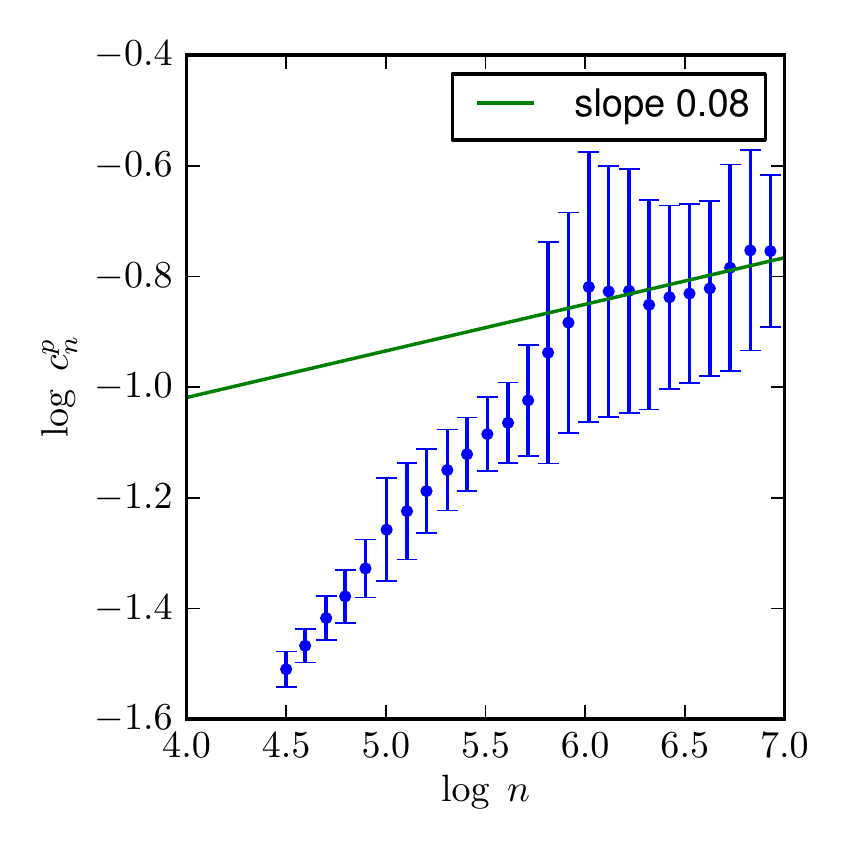}}%
  \subfigure[The symmetric ISAT model, $\tau_c = \tau_x$]{\includegraphics[scale=0.9]{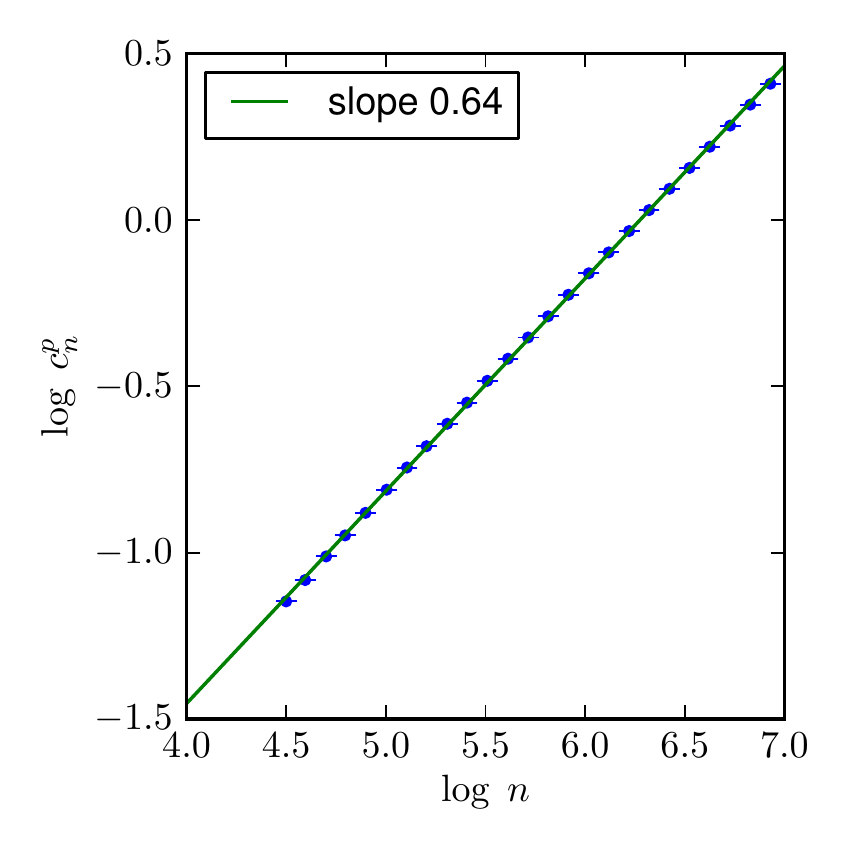}}
  \subfigure[The {\it colliding} model, $\tau_x = 1$]{\includegraphics[scale=0.9]{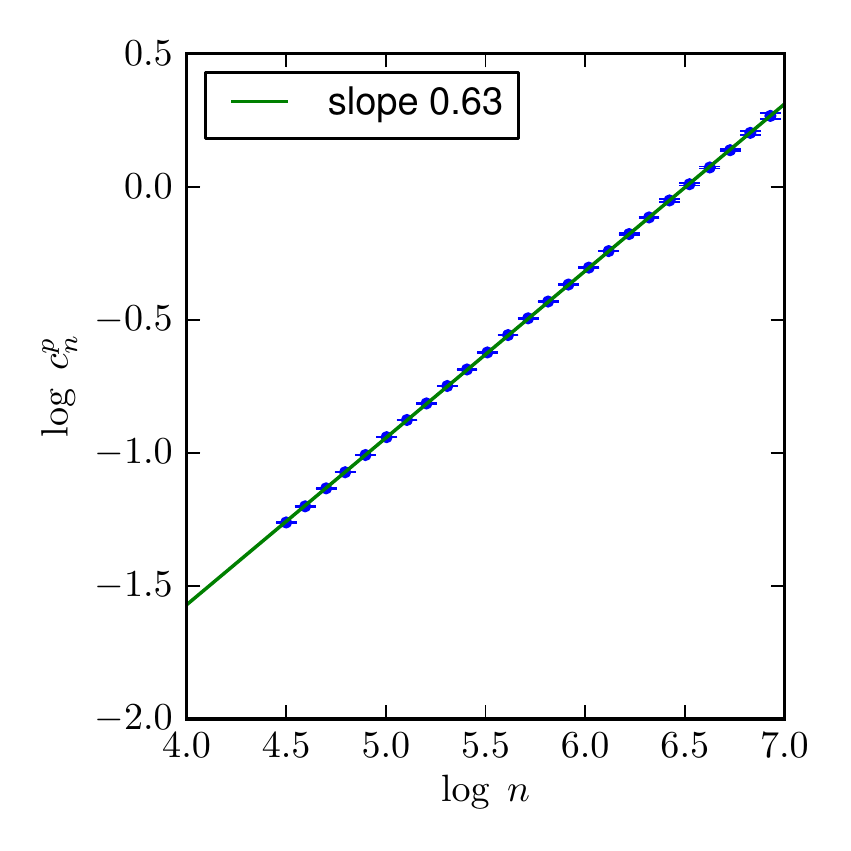}}%
  \subfigure[The VISAW model, $\tau_x = 0$]{\includegraphics[scale=0.9]{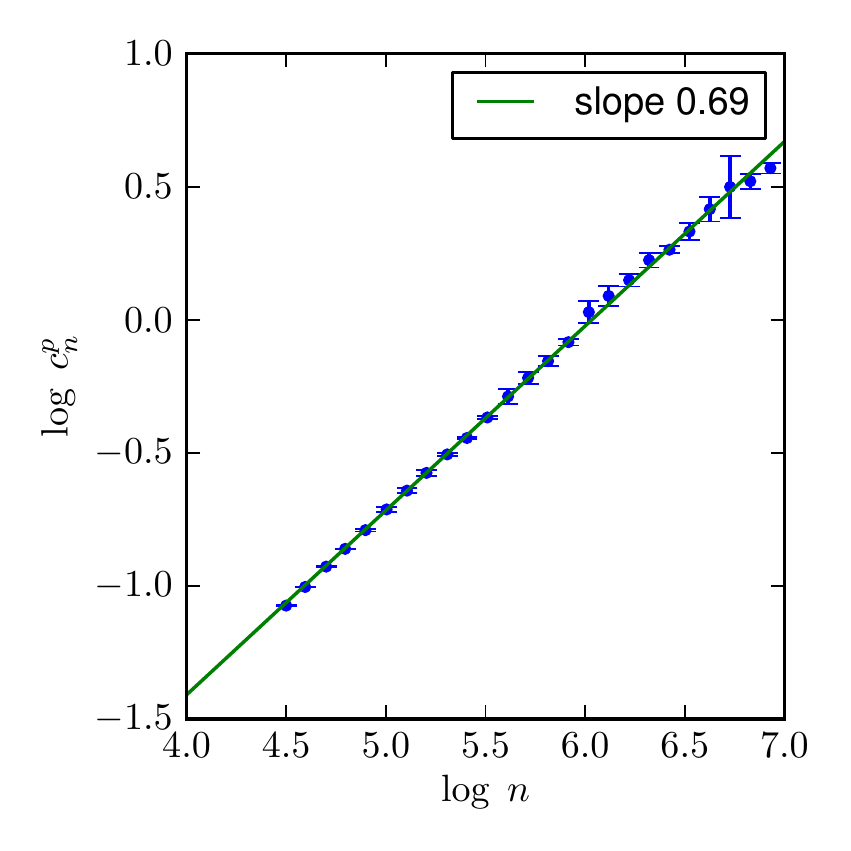}}
  \caption{Double-logarithmic plots of the peak value of the specific heat against length for the four models
defined at the beginning of Section 4. For three of the models the specific heat diverges strongly, while for
the Crossing model the behaviour of the specific heat is markedly different: the data is consistent with a converging 
specific heat.}
  \label{fig:specific-heat}
\end{figure*}

We have begun by analysing the scaling of the specific heat by
calculating the location of its peak $\tau_n^p = \argmax_{\tau}\
c_n(\tau)$ and thereby evaluating $c_n^p = c_n(\tau^p_n)$. In
Fig.~\ref{fig:specific-heat}, we plot the peak values of the specific heat 
for the four models we have simulated. The exponent associated with the peak of the
specific heat, see equation~(\ref{spec-heat-peak-scaling}), is
$\alpha\phi$ if the transition is second order.  For the symmetric ISAT model,
we estimated $\alpha\phi\approx 0.64$, which is a little less
that our previous estimate (equation~(\ref{isat-exponent})) based upon
much longer length trails. We find for the VISAW ($\alpha\phi\approx
0.69$) and the \emph{colliding} model ($\alpha\phi\approx 0.63$)
exponent estimates compatible with that of symmetric ISAT collapse at
this length. This is a crucial result as it confirms the prediction of
Foster \cite{foster2011a-a} that the universality class of ISAT
extends from $r=1/2$ to other values of $r$.

We see that an attempted estimate of $\alpha\phi$ for
the \emph{crossing} model over the full range of lengths is
compromised by poor simulation results. Nevertheless there is clear
curvature in the data and the estimate decreases with length. Using
data from around length $400$ onwards gives an estimate of 0.08. Given
that this estimate would most likely decrease even further with increasing
length, it is likely that the data is compatible with the behaviour of the 
$\theta$-point interacting SAW collapse transition; this has a negative value of the exponent
$\alpha$.

Therefore our data indicates that the ISAT universality
class extends to a range values of the parameter $r$ measuring the 
asymmetry, as conjectured by Foster \cite{foster2011a-a}. 
Intriguingly, rather than becoming first order for $r$ near one as
also conjectured in \cite{foster2011a-a}, the transition becomes weaker 
and potentially $\theta$-like as $r$ increases towards one.

\subsection{Study of the low temperature region}
\label{sec:low-temperature}

To further investigate the proposition that changing the asymmetry
over a range around $r$ does not affect the nature of the collapse,
except potentially near $r$ near 1, we examine the low temperature
phase. This approach proved fruitful in the case of triangular lattice 
extended ISAT \cite{doukas2010a-:a}  where the low temperature phase could be either globular
or fully dense. The fully dense low temperature phase seems associated
with either the ISAT universality class or a first-order transition,
while the globular phase is associated with the much weaker
(non-divergent specific heat) ISAW ($\theta$-point) collapse
transition universality class. Here we present evidence that for the
square lattice symmetric ISAT, \emph{colliding} and VISAW models the
low-temperature phase is maximally dense, and that the density jumps
discontinuously at the critical point. On the other hand for the
\emph{crossing} model the low temperature phase seems not to be fully
dense. This would be compatible with the conjecture that the collapse
transition for the \emph{crossing} model is $\theta$-point-like from a
swollen phase at high temperatures to a globular phase at low
temperatures.

We have considered two different approaches to measuring the
density. The first is indirect by measuring the proportion of sites of
the lattice visited only once by the trail, and the second is using the
radius of gyration to estimate the internal density of the
polymer. Both lead to the same conclusions.

\subsubsection{Proportion of singly visited sites $p_n$}
\label{sec:density_p}

Following the analysis in \cite{doukas2010a-:a} we first measured the
proportion $p_n$ of sites of the lattice visited only once by the
trail. This provides a useful method for considering how dense
our configurations are on average since an asymptotic value of zero
would imply that effectively all the sites occupied by the trail are
doubly occupied. Double occupation of lattice sites implies that the
surrounding edges of the lattice are all occupied with bonds of the
trail: hence the lattice is filled by the trail as the trail
increases in length. At high temperatures it is easy to see that $p_n$
approaches a finite strictly positive value in the thermodynamic
limit: this is connected with the swollen nature of the polymer as
seen in the radius of gyration scaling. One would expect on physical
grounds that $p_\infty(T)$ would be a monotonically increasing
function of temperature $T$.  The question that arises is whether at
low temperatures this value is zero and what kind of singularity
occurs in $p_\infty(T)$ at the collapse transition.

\begin{figure*}[t!]
  \centering
  \subfigure[Critical temperature]{\includegraphics[scale=0.9]{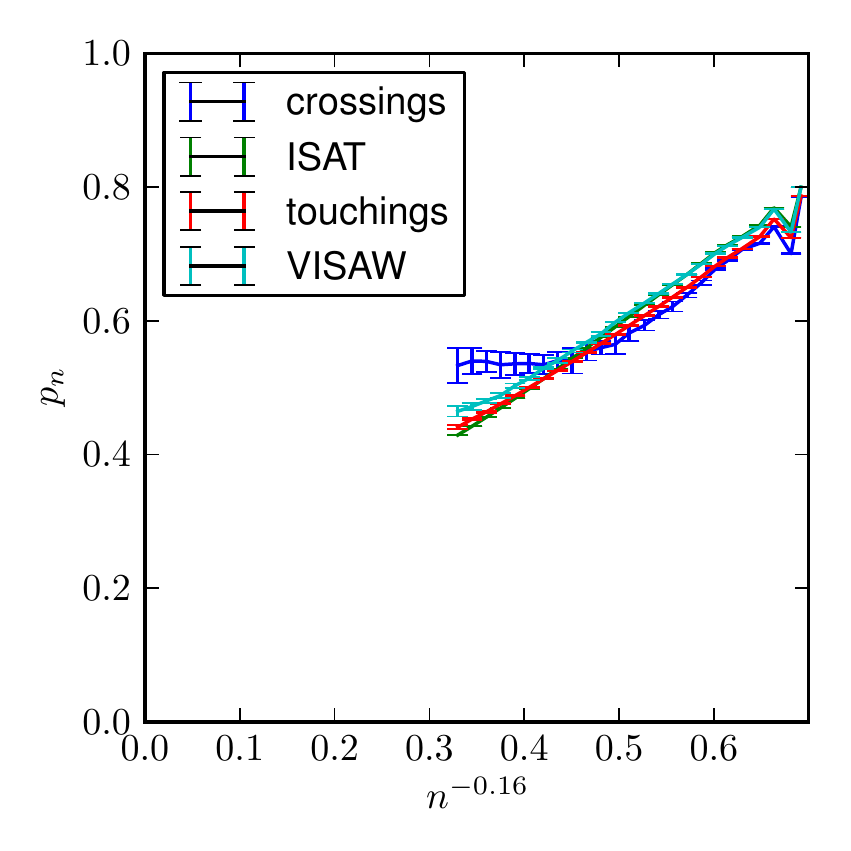}}%
  \subfigure[Low temperature]{\includegraphics[scale=0.9]{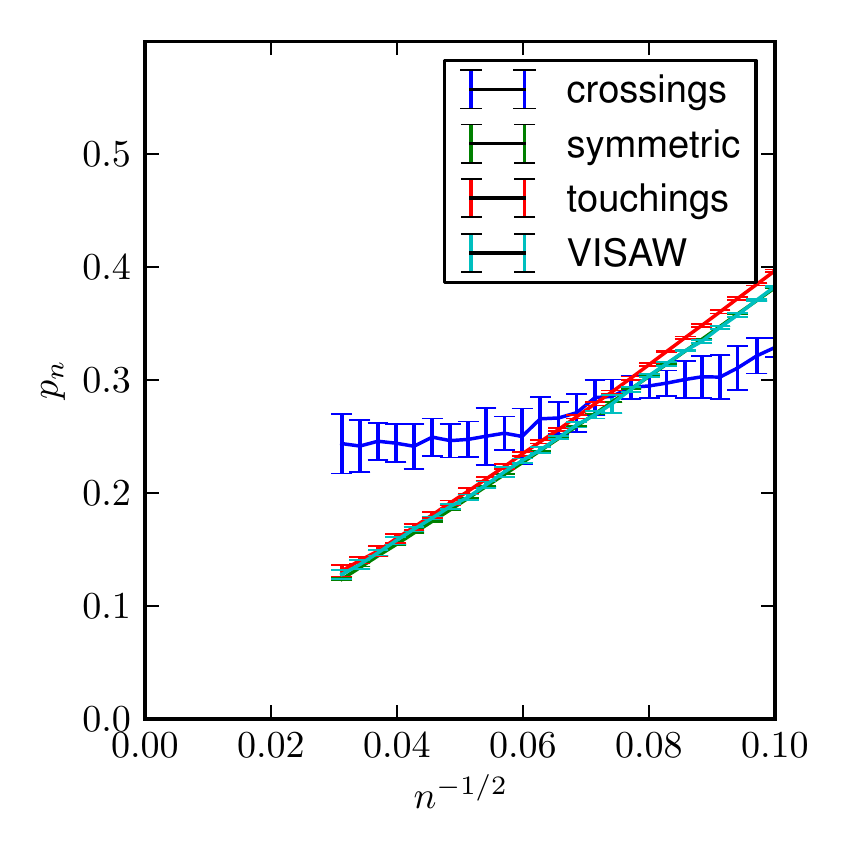}}
  \caption{Plots of $p_n$, the proportion of steps visiting the same
    site once, at a low temperature (on the left) and at the critical
    temperature (on the right). The scale $n^{-1/2}$ chosen is the
    natural low temperature scale. The chosen temperatures for the
    plot on the left are $\tau = 20$ (crossings), $\tau = 5$ (ISAT),
    $\tau = 7$ (colliding), $\tau = 8$ (VISAW). At the critical
    temperature, the quantity $p_n$ is plotted against the crossover
    exponent obtained in \cite{owczarek2007c-:a} $(1-\alpha)\phi
    \simeq 0.16$.}
  \label{fig:density_p}
\end{figure*}

It is worth considering first what happens at the collapse point
itself in the symmetric ISAT model. Thanks to the mapping between
critical ISAT and Kinetic Growth Trails we know that critical limiting
value for $p_n$ is exactly 1/5 \cite{owczarek2007c-:a}. The argument
goes as follows: consider the case when a trail has formed a large
$n$-step loop which occupies $m_1 + m_2 = M$ lattice sites (this is
always the case as trails do not trap). Any site of this loop could
have been the starting point. In order for this site to be visited
only once, the loop must have closed at the first return visit, which
occurs with a probability of $1/3$. Therefore we find for large loops
$m_1/M \to 1/3$, from whence it follows $p_n = m_1/n \to 1/5$. In
Fig.~\ref{fig:density_p} (a), we plotted $p_n$ against $n^{-0.16}$
at the estimated critical temperatures of our four sub-models.  We
have used $(1-\alpha)\phi \simeq 0.16$ as the appropriate correction
to scaling exponent from the estimated exponents for the symmetric
ISAT model with $\alpha\phi \approx 0.68$. For the symmetric ISAT,
\emph{colliding} and VISAW model this choice seems appropriate; in
each case the estimated value of $p_\infty$ is close to $0.2$. While
the correction to scaling exponent may not be appropriate for the
\emph{crossing} model it is clear that any estimate of $p_\infty$ is
greater than $0.2$.

For low temperatures, as discussed above, if the trail fills the
lattice asymptotically in a fully dense phase the portion of monomers
of the trail not involved with doubly-visited sites of the lattice
should tend to zero as $n \to \infty$. In Fig.~\ref{fig:density_p}
(b), we plotted $p_n$ against $n^{-1/2}$ for our four sub-models. The
plots suggest that for the symmetric ISAT, \emph{colliding} and VISAW
model, $p_n \to 0$ as $n \to \infty$ in the low-temperature region,
which implies a maximally-dense phase: our extrapolation estimates for
$p_\infty$ have error bars encompassing zero.  The \emph{crossing}
model seems to show a different behavior: our data suggest that $p_n$
tends to a non-zero value (around $0.25$)

\begin{figure*}[t!]
  \centering
  \subfigure[The {\it crossing} model, $\tau_c = 1$]{\includegraphics[scale=0.9]{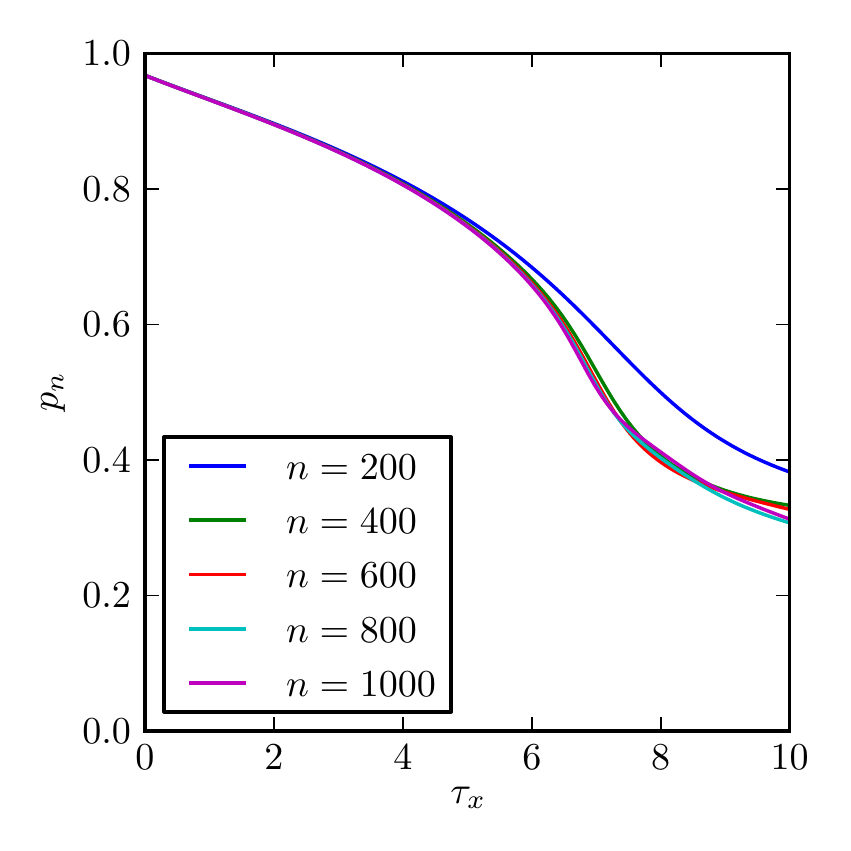}}%
  \subfigure[The symmetric ISAT model, $\tau_c = \tau_x$]{\includegraphics[scale=0.9]{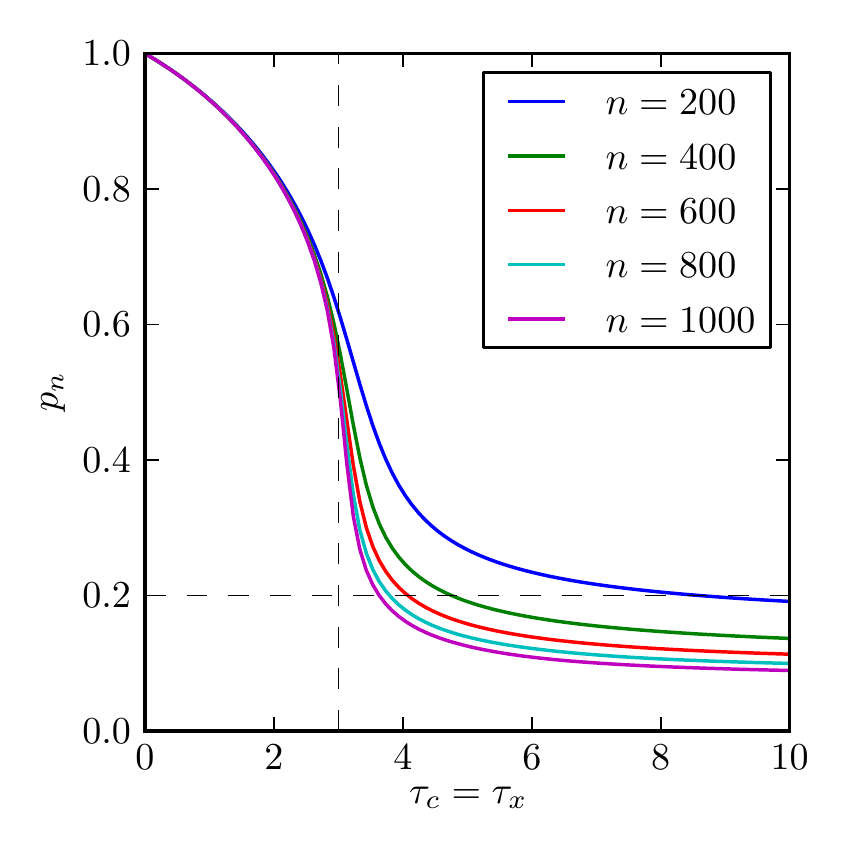}}
  \subfigure[The {\it colliding} model, $\tau_x = 1$]{\includegraphics[scale=0.9]{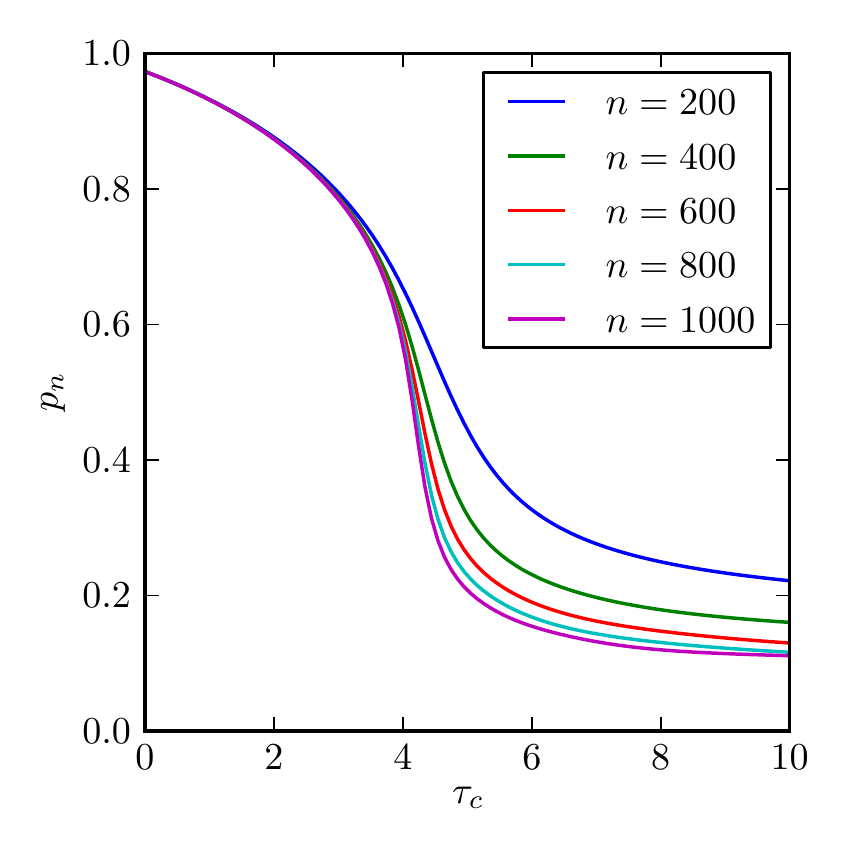}}%
  \subfigure[The VISAW model, $\tau_x = 0$]{\includegraphics[scale=0.9]{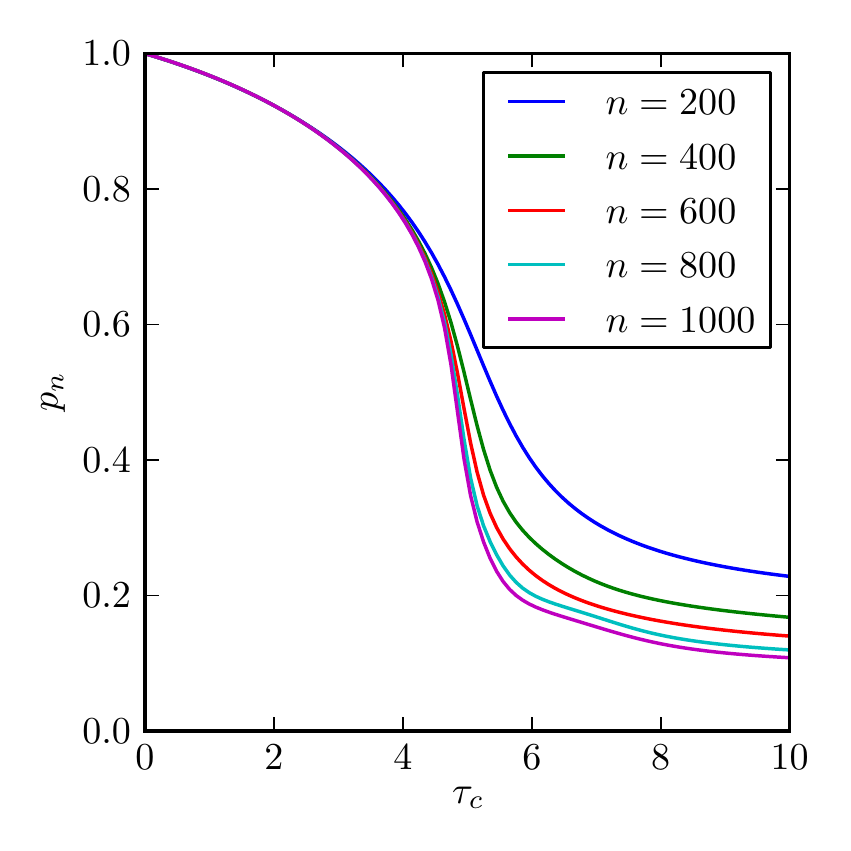}}
  \caption{Density measured as the fraction $p_n$ of sites visited
    only once as function of the length scale $n$ and of the
    temperature, for the four models defined at the beginning of Section 4. 
    In (b) the dashed lines indicate the critical temperature and the
    corresponding value of $p_n$. For three of the models $p_n$ tends to zero above a critical
    value of $\tau$, while for the Crossing model $p_n$ converges to a non-zero value for all
    values of $\tau$.}
  \label{fig:density_p_tau}
\end{figure*}

These results form a clear picture in which
the quantity $p_\infty$ jumps discontinuously to zero at the critical temperature 
for the symmetric ISAT, \emph{colliding} and VISAW models. 
In contrast, for the \emph{crossing} model $p_\infty$ remains non-zero for all temperatures. 
In Fig.~\ref{fig:density_p_tau}, we have plotted the quantity $p_n$
at different length scales as a function of the temperature. The plots
again suggest a common discontinuous behavior for the last three
models, while the crossing model seems to show a continuous transition
in line with the extrapolations described in the previous paragraph.

These findings support the result of the specific heat analysis, namely that the universality
classes of the symmetric ISAT, \emph{colliding} and VISAW models are likely to be the same,
while the \emph{crossing} model is clearly in a distinct universality class.

\subsubsection{Density}
\label{sec:density_rho}

A more direct way of measuring of the density is to consider the
quantity $\rho = n/R^2$ where $R$ is the radius of gyration of the
polymer.

For any AISAT in the high temperature phase we expect $R^2_n \sim
n^{2\nu}$ with $\nu = 3/4$, as for SAWs and SAT, and therefore $\rho_n
= n/R^2_n \to 0$. From \cite{owczarek1995a-:a} we know that at
critical point of the symmetric model $R_n \sim n^{1/2} \log\ n$, so
the density $\rho$ is zero also at the critical point.  For all our
models in the collapsed phase we expect that
$\rho_\infty(T)$ is non-zero at low temperatures. The natural question that arises is
whether $\rho_\infty(T)$ increases from zero at $T=T_c$ as the
temperature is lowered in a continuous fashion, or whether it jumps
discontinuously to a fixed maximum value at soon as the temperature is
smaller than $T_c$, being then constant for all $T<T_c$.

If we take our results for $p_n$ as a guide we would expect that for
the ISAT, the \emph{colliding} and VISAW models $\rho_\infty(T)$ the
density would jump discontinuously on decreasing $T$ below $T_c$ while
for the \emph{crossing} model $\rho_\infty(T)$ would increases
continuously from zero as $T$ is lowered through $T_c$.

\begin{figure*}[t!]
  \centering
  \subfigure[The {\it crossing} model, $\tau_c = 1$]{\includegraphics[scale=0.9]{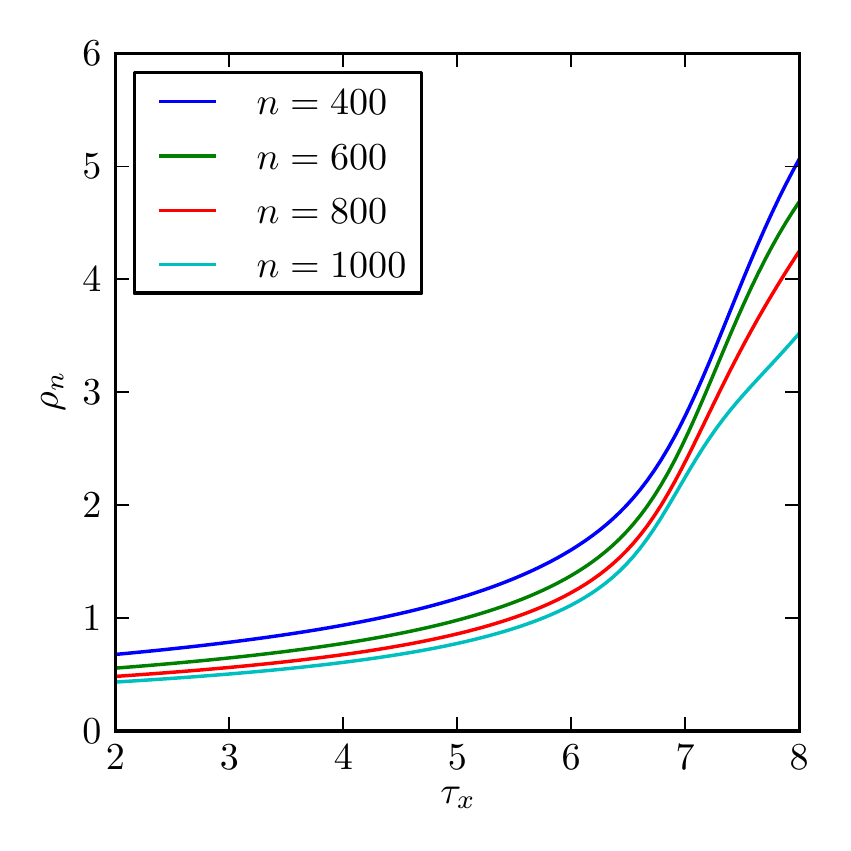}}%
  \subfigure[The symmetric ISAT model, $\tau_c = \tau_x$]{\includegraphics[scale=0.9]{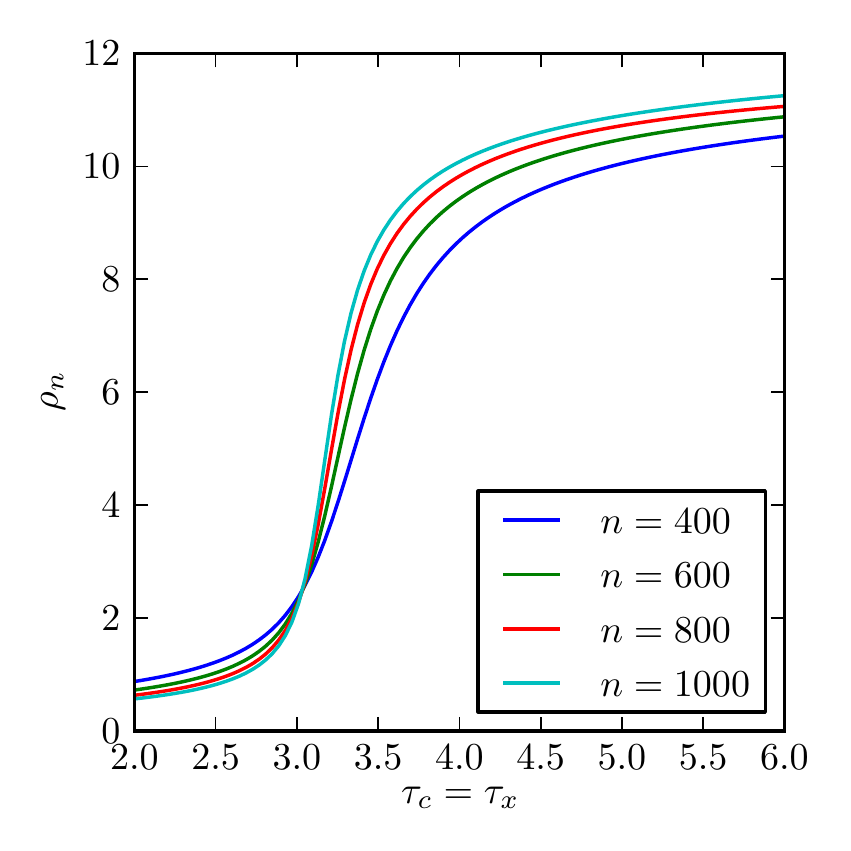}}
  \subfigure[The {\it colliding} model, $\tau_x = 1$]{\includegraphics[scale=0.9]{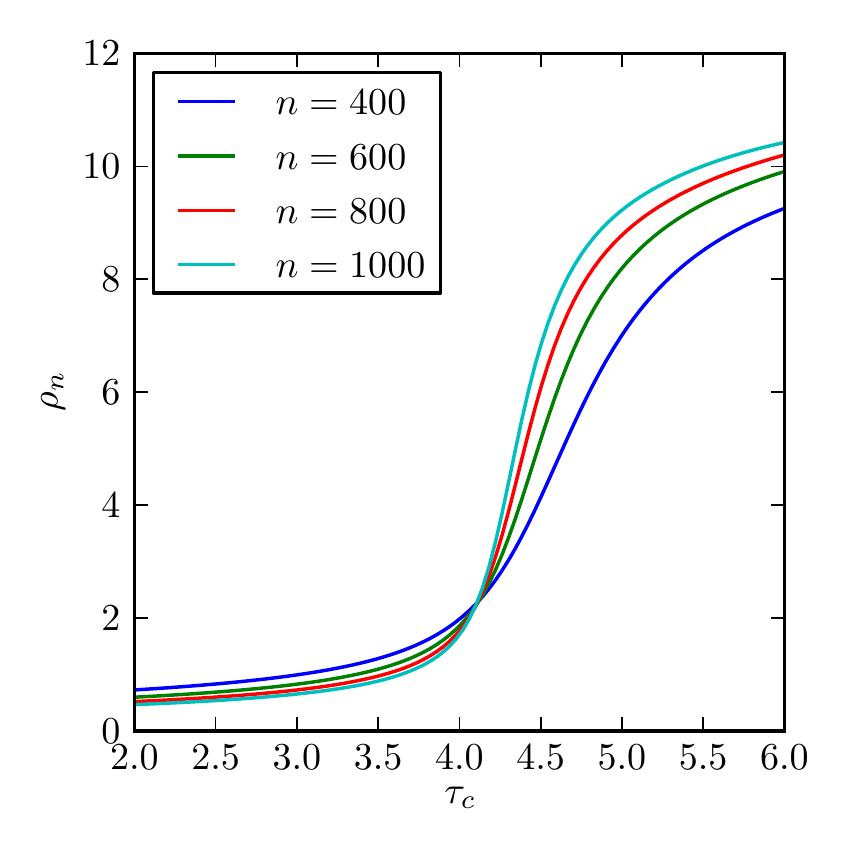}}%
  \subfigure[The VISAW model, $\tau_x = 0$]{\includegraphics[scale=0.9]{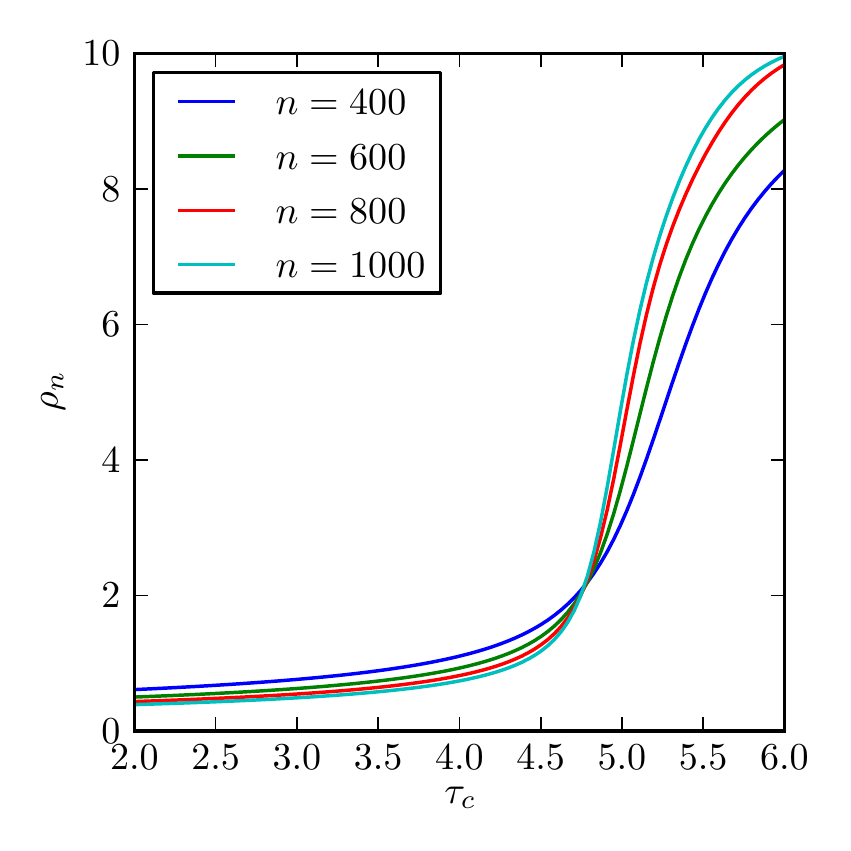}}
  \caption{Plot of the density $\rho_n = n/R_n^2$ as function of the temperature,
    for the four models defined at the beginning of Section 4. 
    For three of the models there is a clear crossing point such that for small values
    of $\tau$ the value of the density decreases with length, while for large values of
    $\tau$ the value of the density increases with length. For the Crossing model no
    such crossing point exists, and the density decreases with length for all values of $\tau$.}
  \label{fig:density_rho}
\end{figure*}

In Fig.~\ref{fig:density_rho} we plotted $\rho_n = n / R_n^2$ as a
function of the temperature. In the symmetric ISAT, \emph{colliding}
and VISAW models the density plotted for different lengths cross at a
Boltzmann weight close to the expected collapse transition point,
increasing to the right and decreasing to the left. As argued above,
the curves converge to zero for small Boltzmann weights (high
temperatures) and are consistent with an increase to a temperature-independent
constant for large Boltzmann weights (low temperatures). This is consistent with
the behaviour of $p_n$ discussed previously. The \emph{crossing}
model seems to have a different behavior. Consistent with a
varying limiting value of $p_n$, the limiting value of the density
$\rho_n$ does not approach a temperature-independent constant, either.

\section{Conclusions}
\label{sec:conclusions}

We have studied a generalised model of interacting self-avoiding
trails (ISAT) on the square lattice as proposed by Foster
\cite{foster2011a-a} where the weight associated with crossing-type
interactions ($\tau_x$) and collision-type interactions ($\tau_c$) may
differ.

From our analysis, we can confirm the conjecture in
\cite{foster2011a-a} that the ISAT universality class extends over a
region of asymmetry around $\tau_x=\tau_c$. We can conclude that this
region extends down to $\tau_x=0$, which is also known as the VISAW
model, and seems extends to some larger $\tau_x > \tau_c$. Our
simulation results for $\tau_c=1$ are compromised numerically by
poor convergence: in fact at small lengths we see some evidence of
multiply peaked probability distributions but these seem to become
unimodal at larger lengths. Importantly, the peak of the specific heat
seems to decrease on increasing $x=\tau_x/\tau_c$. Indeed the specific
heat of the $\tau_c=1$ model diverges with an effective exponent that
is much smaller (albeit with large error bars) than the ISAT
universality class would predict and, in fact, the specific heat may
not diverge: we would expect a more strongly diverging specific heat
if a first-order transition occurs. Supporting these conclusions is
our investigation of the low temperature phase for different
asymmetry. For the symmetric ISAT and the VISAW model the low
temperature phase seems to be fully dense as it is in the extended
triangular lattice model in certain regimes. Also, in agreement with
our tentative prediction for the $\tau_c=1$ case is the evidence that
here the low temperature phase is no longer fully dense, which implies
a globular ISAW-like low temperature phase. Putting this information
together leads us to predict a $\theta$-point-like collapse as occurs
in interacting self-avoiding walks for $x$ sufficiently large. Further
numerical work is clearly needed to pin down the large $x$ behaviour
of the AISAT model, including where the change of universality class occurs. 
Finally, we point out that our evidence that the
low temperature phase is fully dense for symmetric ISAT implies that
there is a first-order characteristic of this transition, as predicted by Foster
\cite{foster2011a-a}, even though there is no
latent heat.

\section*{Acknowledgements}

Financial support from the Australian Research Council via its support
for the Centre of Excellence for Mathematics and Statistics of Complex
Systems and through its Discovery program is gratefully acknowledged
by the authors. A L Owczarek thanks the School of Mathematical
Sciences, Queen Mary, University of London for hospitality.


\end{document}